\documentclass[lettersize,journal]{IEEEtran}
\usepackage{amsmath,amsfonts}
\usepackage{algorithmic}
\usepackage{array}
\usepackage[caption=false,font=normalsize,labelfont=sf,textfont=sf]{subfig}
\usepackage{textcomp}
\usepackage{stfloats}
\usepackage{url}
\usepackage{verbatim}
\usepackage{graphicx}
\def\BibTeX{{\rm B\kern-.05em{\sc i\kern-.025em b}\kern-.08em
    T\kern-.1667em\lower.7ex\hbox{E}\kern-.125emX}}
\usepackage{balance}
\begin{document}
\title{DiSC: Resolution-Scalable Acceleration of Diffusion Models by Exploiting Sparsity and Cached Token Reuse with Hash-based Distribution}
\author{
Jieon Yoon,~\IEEEmembership{Graduate Student Member,~IEEE}, 
Hangyeol Lee,~\IEEEmembership{Graduate Student Member,~IEEE},
Jaehoon Heo,~\IEEEmembership{Graduate Student Member,~IEEE},
and Joo-Young Kim,~\IEEEmembership{Senior Member,~IEEE}
\thanks{This work was supported in part by the Institute of Information \& Communications Technology Planning \& Evaluation (IITP) under the Graduate School of Artificial Intelligence Semiconductor Program, funded by the Korea government (MSIT), under Grant IITP-2026-RS-2023-00256472; and in part by the IITP-Information Technology Research Center (IITP-ITRC) Program, funded by the Korea government (MSIT), under Grant IITP-2026-RS-2020-II201847. The EDA tool was supported by the IC Design Education Center (IDEC), Korea.}
\thanks{Jieon Yoon is with the Graduate School of AI Semiconductor, Korea
Advanced Institute of Science and Technology (KAIST), Daejeon 34141,
South Korea (e-mail: j9e8y@kaist.ac.kr).}
\thanks{
Hangyeol Lee, Jaehoon Heo, and Joo-Young Kim are with the Department of Electrical Engineering, Korea
Advanced Institute of Science and Technology (KAIST), Daejeon 34141,
South Korea (e-mail: lhg4294@kaist.ac.kr; kd01050@kaist.ac.kr; jooyoung1203@kaist.ac.kr).
}
}


\maketitle

\begin{abstract}
Transformer-based diffusion models offer superior scalability and performance but suffer from high computational overhead due to the iterative nature and quadratic complexity of self-attention at high resolutions.
In this paper, we propose DiSC, a resolution-scalable, sparsity-aware hardware accelerator.
At the software level, DiSC introduces two algorithms: Cached Token Reuse (CTR), and Softmax Thresholding with Sparsity Mask Reuse (ST).
CTR introduces a mechanism that translates spatial variations in the input latent difference across steps into a token-level reuse decision, effectively eliminating redundant token computation.
ST induces sparsity in attention operations by reusing a generated sparsity pattern, leveraging temporal similarity to bypass costly prediction overhead.
Together, these algorithms provide resolution-scalable computational benefits and yield a moderate sparsity and hybrid dense-sparse workload.

To exploit this efficiently, we design a specialized hardware architecture and unified dataflow.
This architecture avoids dedicated sparsity-handling components; instead, a hash–based distribution over on-chip memory banks allows DiSC to reuse its existing compute engines for sparse operations, efficiently exploiting the induced sparsity with minimal hardware overhead.
Evaluated on DiT and PixArt-$\Sigma$, DiSC achieves 3.47–4.74$\times$ and 2.48–3.50$\times$ speedups over NVIDIA A100 and H100 GPUs, respectively, with energy savings ranging from 46.4\% to 68.1\%.
\end{abstract}


\section{Introduction}
\label{sec:introduction}
\IEEEPARstart{R}{ecently}, diffusion models have attracted significant attention for high-quality image generation capabilities. Based on the model architecture, diffusion models can be classified into models with U-Net backbone and models with transformer backbone. Earlier diffusion models, such as Stable Diffusion~\cite{rombach2022high} and Guided Diffusion~\cite{dhariwal2021diffusion}, adopted a U-Net backbone. However, the introduction of DiT~\cite{peebles2023scalable} demonstrated that transformer-based diffusion models can outperform U-Net-based models in terms of scalability and efficiency. Following this trend, many subsequent models, such as Stable Diffusion3~\cite{esser2024scalingrectifiedflowtransformers}, PixArt-$\alpha$~\cite{chen2023pixart} and PixArt-$\Sigma$~\cite{chen2024pixart} have adopted transformer backbone.

Due to their inherent iterative nature---requiring multiple denoising steps---diffusion models incur significantly higher computational cost relative their model size. 
This computational burden is further exacerbated by the growing demand for high-resolution generation.
Consequently, recent transformer-based diffusion models have been designed to support increasingly larger resolutions: DiT can generate images up to 512$\times$512, PixArt-$\alpha$ up to 1024$\times$1024, and PixArt-$\Sigma$ targets resolutions up to 4K (with publicly released weights available up to 2K).

\begin{figure}[t]
  \centering
  \includegraphics[width=\linewidth]{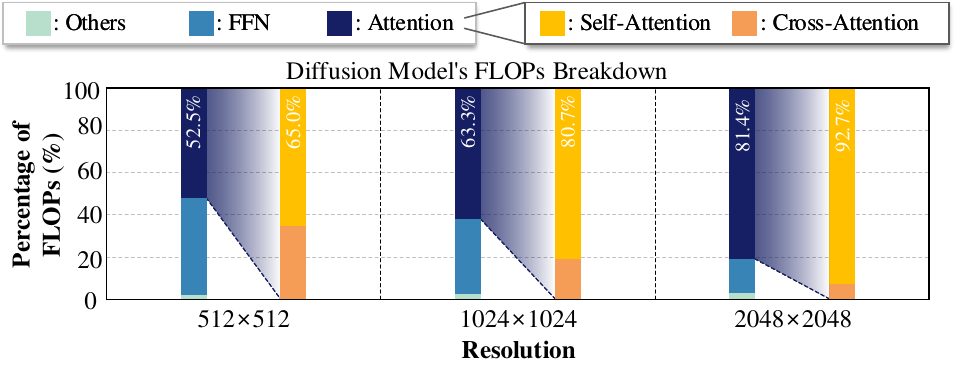}
  \caption{FLOPs Breakdown of PixArt-$\Sigma$ across different resolutions}
  \label{fig1}
\end{figure}

In contrast to U-Net-based models, which are composed of residual blocks, convolutional blocks, and transformer blocks, transformer-based diffusion models consist solely of transformer blocks. This architectural shift concentrates the computational workload in the attention layers, and the effect becomes particularly severe at high resolutions. As shown in Figure~\ref{fig1}, when generating a 2048$\times$2048 image using PixArt-$\Sigma$, self-attention operations account for over 75\% of the total computation.
Since self-attention has quadratic complexity, $O(N^2)$, where the token length $N$ grows with image resolution, attention dominates the end-to-end computation as resolution increases.

To mitigate the computational challenge of diffusion models, many training-free acceleration studies have been proposed.
Most of them exploit the inherent temporal similarity of the diffusion models, often by reusing features from adjacent timesteps through caching-based schemes~\cite{selvaraju2024fora,ma2024learning,ma2024deepcache,zou2024accelerating,zou2024DuCa, chen2025icc}. 
Recent works apply selective feature reuse to different components—DiTFastAttn~\cite{yuan2024ditfastattn} adopts a coarse-grained reuse strategy in the attention layers, whereas EXION~\cite{heo2025exionexploitinginterintraiteration} introduces a fine-grained reuse scheme in the FFN layers along with dedicated hardware support.
Furthermore, Cambricon-D~\cite{kong2024cambricon} and Ditto~\cite{kim2025ditto} leverage differential computation across adjacent timesteps, applying quantization to inter-step differences and introducing dedicated hardware to efficiently process them.

From these prior works, two key observations can be made.
First, there has been little research targeting high-resolution cases in transformer-based diffusion models.
Methods such as ~\cite{selvaraju2024fora, ma2024deepcache, kong2024cambricon} address only U-Net-based diffusion models, while~\cite{zou2024accelerating, zou2024DuCa, heo2025exionexploitinginterintraiteration, kim2025ditto} target transformer-based models but evaluate only at low resolution (e.g., 256$\times$256).
Second, in the domain of transformer-based diffusion models, unlike LLMs or vision transformers, where some tokens can be treated as low-importance and safely approximated or skipped~\cite{ham20203, wang2021spatten, ham2021elsa, qin2023fact}, image-generation diffusion models must generate all tokens, because each token corresponds to a specific spatial region in the output image. Aggressively skipping tokens often leads to visible local artifacts rather than small, distributed errors. 

These observations highlight the need for new acceleration methods that remain effective for high-resolution transformer-based diffusion models, and that specifically target the rapidly growing cost of attention while minimizing quality degradation.

To this end, we propose DiSC, a resolution-scalable hardware accelerator for transformer-based diffusion models, effective across both low- and high-resolution settings.
DiSC jointly leverages inherent temporal similarity of diffusion models and the input/output sparsity that emerges in attention operations, integrating software-level algorithms with a specialized hardware architecture that maximizes their effectiveness.

On the software side, DiSC introduces two complementary training-free methods: Cached Token Reuse (CTR) and Softmax Thresholding with Sparsity Mask Reuse (ST). 
CTR exploits the temporal similarity across denoising steps to reduce redundant token computation in both the attention and feed-forward network (FFN) layers through a lightweight token-selection mechanism leveraging the one-to-one correspondence between latent patches and tokens.
ST reduces attention computation by inducing input and output sparsity: it sparsifies the attention probability matrix by thresholding negligible probabilities and reuses the resulting mask across nearby denoising steps with minimal accuracy degradation.
CTR and ST are complementary across resolutions. In our evaluations, CTR contributes more at lower resolutions, where a larger token pruning ratio is achievable with minimal impact on image quality, whereas at higher resolutions, where the $O(N^2)$ attention cost dominates, ST delivers larger FLOPs reductions. By combining the two, DiSC achieves resolution-scalable acceleration from low to high resolutions.

However, existing hardware architectures become suboptimal for the specific workload induced by these algorithms.
The primary reason is that the sparsity induced by ST is moderate (e.g., 56\% in PixArt-$\Sigma$), unlike the high sparsity (e.g., $>$99\%) targeted by prior sparse accelerators.
Furthermore, given that layers such as FFNs remain dense, applying traditional sparse accelerators to this moderate sparsity and hybrid dense-sparse workload results in an inefficient area trade-off, as their dedicated sparse units suffer from severe under-utilization during the dense operations.
Therefore, a new architecture is required that can efficiently handle this specific workload.

At the hardware level, we propose a specialized architecture and a unified dataflow that are purpose-built for this workload.
Instead of adding dedicated sparse hardware, our architecture repurposes its existing compute units to handle the sparse operations.
A hash–based distribution maps work across on-chip SRAM banks so that each compute unit accesses only its own bank, avoiding conflicts and costly inter-bank routing while fully exploiting independent bank access.
With an appropriate choice of hash function, this mapping also balances non-zero work across units, which is crucial to actually converting sparsity into performance gains.

The contributions of this paper are summarized as follows:
\begin{itemize}
    \item We present DiSC, a training-free, resolution-scalable HW–SW co-designed accelerator for transformer-based diffusion models. 
    \item We introduce two complementary training-free methods—Cached Token Reuse (CTR) and Softmax Thresholding with Sparsity Mask Reuse (ST)—that exploit temporal similarity and attention sparsity across timesteps. Applied jointly, they provide resolution-scalable acceleration from low to high resolutions.
    \item We design a specialized hardware architecture and unified dataflow tailored to the moderate sparsity and hybrid dense–sparse workloads induced by CTR and ST. A hash–based distribution over on-chip memory banks allows DiSC to reuse its existing compute engines for sparse operations, eliminating the need for dedicated sparse hardware.
    \item DiSC achieves 3.47-4.74$\times$ speedup and 61.3-68.1\% energy reduction over an NVIDIA A100 GPU, and 2.48-3.5$\times$ speedup and 46.4-62.1\% energy reduction over an NVIDIA H100.
\end{itemize}
\section{Background and Motivation}
\subsection{Overview of the Diffusion Model}
\begin{figure}[t]
  \centering
  \includegraphics[width=\linewidth]{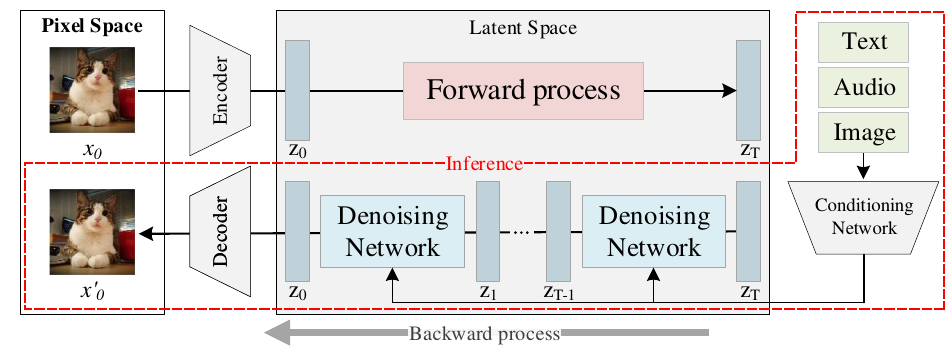}
  \caption{Overview of Diffusion Model}
  \label{fig2}
\end{figure}
\noindent Figure~\ref{fig2} illustrates the overview of the diffusion model, which consists of a forward process and a backward (denoising) process. 
As the term "diffusion" implies, these processes involve the gradual addition and removal of noise, respectively, over multiple steps. 
During the forward process, noise is progressively added to $z_0$, a latent representation of original image $x_0$ obtained through an encoder, eventually transforming it into the noise latent $z_T$. 
In the backward process, the denoising network iteratively removes the noise from $z_T$, under the guidance of a condition embedding derived from the given input (e.g., text) via a conditioning network. 
During training, the model alternates between the forward and backward processes to learn a denoising network. 
During inference, only the backward process is executed, starting from random noise $z_T$ to generate an image. In this work, we focus on optimizing the inference process of diffusion models.

\subsection{Transformer-based Diffusion Model}
\begin{figure}[t]
  \centering
  \includegraphics[width=\linewidth]{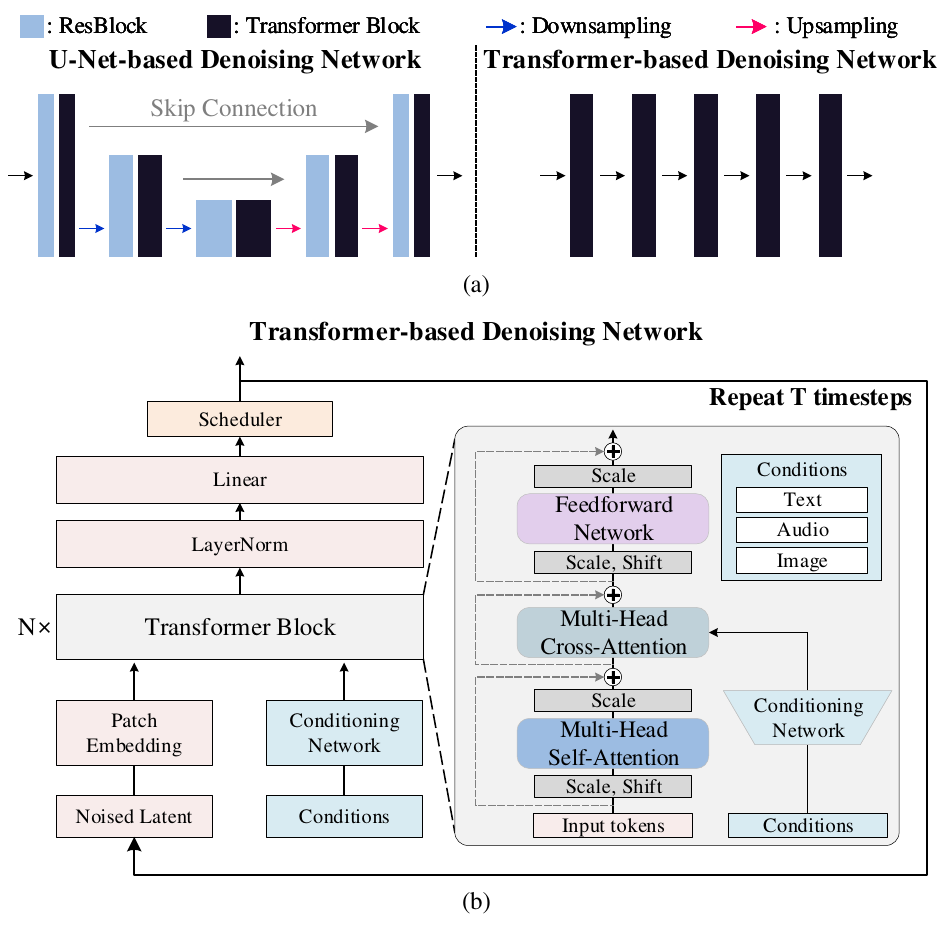}
  \caption{(a) Backbone Type of Denoising Network (b) Model Architecture of Transformer-based Denoising Network for Diffusion Models}
  \label{fig3}
\end{figure}
\noindent Figure~\ref{fig3}(a) illustrates two types of denoising networks. 
The first is a U-Net-based denoising network composed of transformer blocks and residual blocks, along with separate downsampling and upsampling blocks that reduce and then restore the spatial dimension. The second type is a transformer-based denoising network, which consists of a sequence of transformer blocks. 
In this work, we focus on the transformer-based architecture, whose detailed structure is illustrated in Figure~\ref{fig3}(b).
A transformer block---the fundamental building block of the transformer-based denoising network---typically consists of a self-attention layer, a cross-attention layer, and a feed-forward network (FFN). 
Assuming the number of tokens is $N$, the computational complexity of the cross-attention layer and FFN is $O(N)$, while that of self-attention is $O(N^2)$. 
Since $N$ is proportional to the image resolution, this quadratic complexity makes self-attention the dominant computational bottleneck, particularly for high-resolution generation where $N$ becomes large.

\subsection{Motivation for Cached Token Reuse}
\begin{figure}[t]
  \centering
  \includegraphics[width=\linewidth]{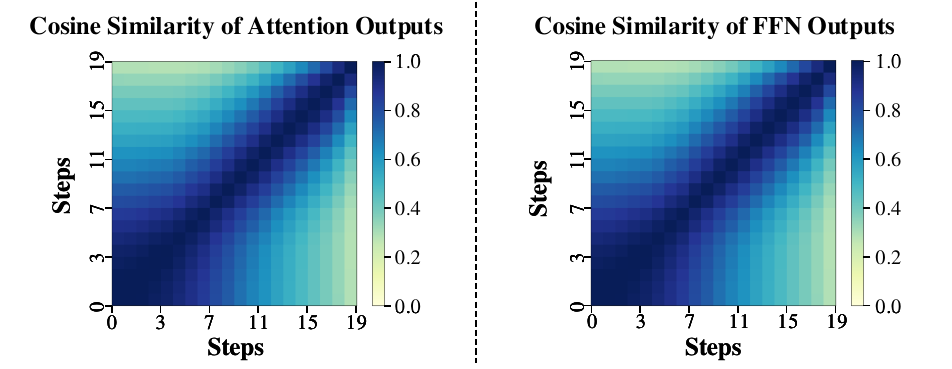}
  \caption{Heatmap of cosine similarity between the outputs of attention and FFN layer}
  \vspace{-0.12in}
  \label{fig5}
\end{figure}

\noindent The iterative denoising process of diffusion models inherently results in a high degree of temporal similarity in intermediate layer outputs across adjacent timesteps.
This characteristic is well-established and has been leveraged by prior approaches~\cite{selvaraju2024fora, ma2024deepcache, ma2024learning, wimbauer2024cache} to reduce computation by reusing layer outputs.
Figure~\ref{fig5} corroborates this behavior in the PixArt-$\Sigma$ model, demonstrating high cosine similarity among the outputs of both attention and FFN layers across adjacent steps.

\begin{figure}[t]
  \centering
  \includegraphics[width=\linewidth]{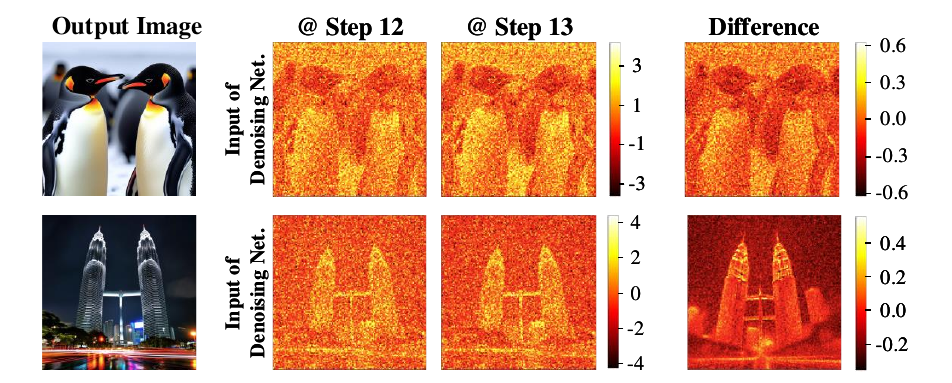}
  \caption{Heatmap of input latent $z_t$ (1st head shown from total 4 heads) at adjacent timesteps (12th and 13th timestep) and their difference in the PixArt-$\Sigma$ model}
  \label{fig6}
\end{figure}

However, Figure~\ref{fig6} reveals a critical observation: the temporal difference in input latent ($z_t$ - $z_{t-1}$) is spatially non-uniform.
While some spatial positions undergo substantial changes, others exhibit negligible differences.
This implies that prior approaches relying on coarse-grained, layer-level caching are inefficient, as they process the entire input indiscriminately.
In contrast, this spatial non-uniformity of input latent difference implies an opportunity to optimize caching methods by explicitly distinguishing between regions that require updates and those that can be reused.
This motivates the need for a more granular, token-level reuse strategy.
\begin{figure}[t]
  \centering
  \includegraphics[width=\linewidth]{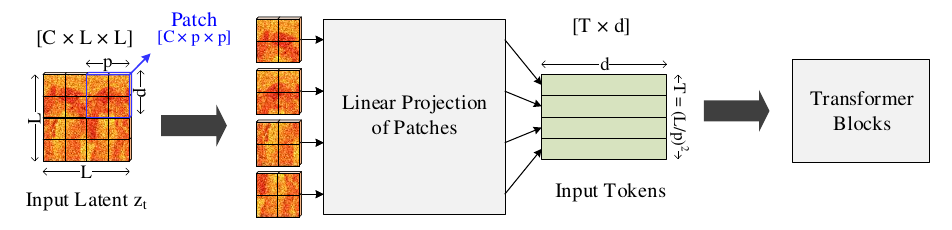}
  \caption{The Process of Patch Embedding}
  \label{fig7}
\end{figure}

To translate this observation into an effective token-level reuse strategy, we leverage the diffusion model's patch embedding process.
Widely adopted in vision transformers (ViT) and diffusion transformers (DiT), patch embedding is the fundamental mechanism that enables transformers to process visual data.
In the transformer-based denoising network, each timestep initiates with this patch embedding layer, followed by a sequence of transformer blocks.

As illustrated in Figure~\ref{fig7}, the patch embedding process partitions the latent input into distinct patches and projects each patch into a corresponding token via linear embedding.
Consequently, each spatial patch in the input latent translates to a single token, and these tokens are fed into the subsequent transformer blocks. 
Leveraging this structural characteristic, we propose a token selection mechanism aligned with the patch embedding process. 
This approach allows us to selectively determine which tokens require recomputation based on latent input variations and which can be reused.
The detailed scheme is described in Section~\ref{sec:ctp}.
\subsection{Motivation for Softmax Thresholding with Sparsity Mask Reuse}
\label{subsec:st_motiv}
\begin{figure}[t]
  \centering
  \includegraphics[width=\linewidth]{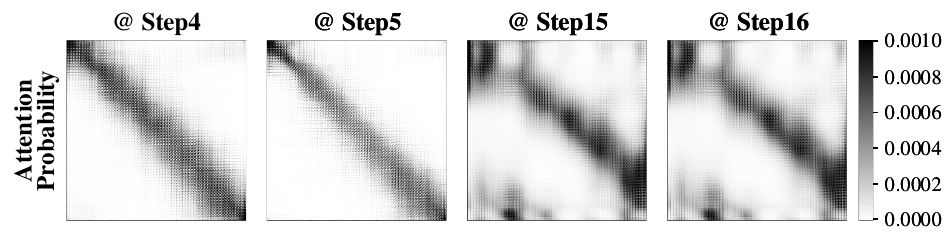}
  \caption{Heatmap of Attention Probability on PixArt-$\Sigma$ at 1k$\times$1k. Values beyond the color scale range are shown in the same color (black) as the maximum value in the range.}
  \label{fig8}
\end{figure}

\noindent While inducing sparsity in attention mechanisms has been explored in other domains~\cite{ you2023vitcod, lu2021sanger, lee2023hammer, qin2023fact}, applying it to diffusion models presents unique opportunities due to the temporal similarity inherent in the diffusion models.
Figure~\ref{fig8} visualizes the attention probabilities in PixArt-$\Sigma$, revealing two key observations: first, a significant portion of values are negligible (near-zero); second, the pattern of these near-zero values exhibits high temporal similarity across adjacent timesteps.

Typically, leveraging input sparsity in the subsequent operation ($P \cdot V$) is straightforward once the sparse attention probabilities $P$ is generated.
However, exploiting sparsity in the preceding operation ($Q \cdot K^T$) is challenging because the sparsity pattern is unknown prior to computation.
To address this, prior works devised prediction mechanisms to induce output sparsity: ViTCoD~\cite{you2023vitcod} enforces fixed sparse patterns, while Sanger~\cite{lu2021sanger}, HAMMER~\cite{lee2023hammer} and FACT~\cite{qin2023fact} perform precomputation using low-bit precision to estimate the sparsity mask.
In contrast, the temporal similarity in diffusion models allows us to 
bypass these prediction steps by simply reusing the mask from 
the adjacent timestep, effectively enabling sparse computation for 
$Q \cdot K^T$ without additional predictive overhead.

However, this approach introduces critical constraints for hardware acceleration.
First, considering the trade-off between sparse attention and generation quality, the achievable sparsity remains moderate (50--60\%) to preserve model accuracy.
Second, this sparsity is limited to the attention layers, while other layers, such as FFNs, remain dense.
This resulting moderate sparsity and hybrid sparse-dense workload presents a challenge not addressed by conventional sparse accelerators.
\subsection{Hardware Motivation under Moderate Sparsity}
\label{subsec::mot_hardware}
\begin{figure}[t]
  \centering
  \includegraphics[width=\linewidth]{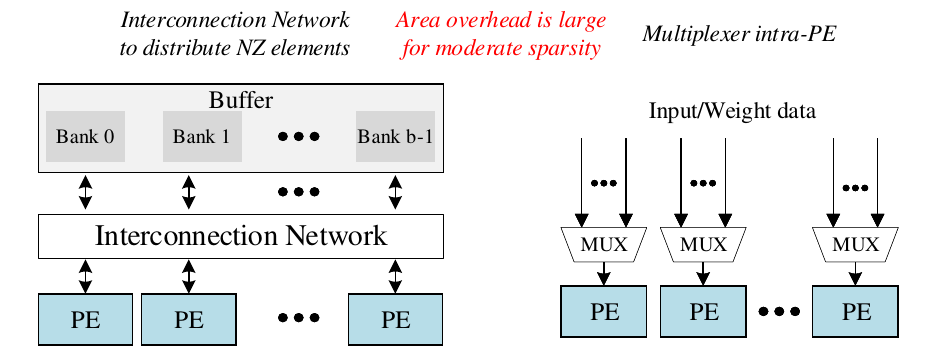}
  \caption{Representative hardware suggestion for sparsity}
  \label{fig_over}
\end{figure}

\noindent The unique workload characteristics---moderate sparsity and a hybrid sparse-dense workload---pose a significant challenge for existing hardware architectures.
It is well-established that while commodity GPUs can handle structured or block sparsity, the irregularity inherent in fine-grained, unstructured patterns is ill-suited for parallel computing systems like GPUs, as it leads to severe load imbalance that significantly limits performance.
To overcome these limitations, various specialized accelerators have been proposed. For instance, one line of research~\cite{srivastava2020matraptor, zhang2021gamma, pal2018outerspace} focuses on SpGEMM (Sparse General Matrix-Matrix Multiplication) accelerators. However, these are typically designed for workloads with extremely high sparsity ($>$99\%) and are not optimized for our hybrid sparse-dense workload.

A more relevant line of work~\cite{lu2021sanger, qin2020sigma, parashar2017scnn, yang2024trapezoid, heo2025exionexploitinginterintraiteration} suggests accelerators capable of handling both dense and sparse operations. These accelerators have followed two architectural directions to exploit sparsity, as illustrated in Figure~\ref{fig_over}. The first~\cite{qin2020sigma, parashar2017scnn, yang2024trapezoid} employs flexible interconnection networks (e.g., crossbars or Benes networks) to dynamically map non-zero operands to PEs. The second direction~\cite{heo2025exionexploitinginterintraiteration, lu2021sanger} moves the flexibility inside each PE by wiring multiple input and weight lines and selecting the desired non-zero operand with multiplexers.
While effective at high sparsity, these approaches incur non‑trivial overhead for our moderate sparsity workload. 
For example, SIGMA’s distribution and reduction networks add 37.7\% chip area over an iso-PE systolic baseline~\cite{qin2020sigma}, and Trapezoid reports that its distribution network occupies more than half the area of the multipliers~\cite{yang2024trapezoid}.
This heavy overhead is also entirely underutilized during the 
dense computations, making these designs inefficient for our 
moderate sparsity and hybrid sparse-dense workload.
\section{DiSC's Algorithmic Optimizations}
\noindent In this section, we present two training-free algorithmic optimizations for transformer-based diffusion models: Cached Token Reuse (CTR) and Softmax Thresholding with Sparsity Mask Reuse (ST). 

CTR introduces a selective, token-level pruning mechanism that moves beyond prior coarse-grained caching methods.
It identifies redundant tokens by spatial variations in the input latent difference between adjacent steps.
ST exploits the prevalence of near-zero values in attention probabilities and leverage the high temporal similarity of these sparsity pattern.
By generating an attention sparsity mask and reusing it across subsequent steps, ST induces sparsity in both the $Q \cdot K^T$ and $P \cdot V$ operations, bypassing the prediction overhead required by other sparse attention techniques~\cite{lu2021sanger, qin2023fact, lee2023hammer} to estimate the $Q \cdot K^T$ sparsity pattern ahead of time.

\subsection{Cached Token Reuse (CTR)}
\label{sec:ctp}
\begin{figure}[t]
  \centering
  \includegraphics[width=\linewidth]{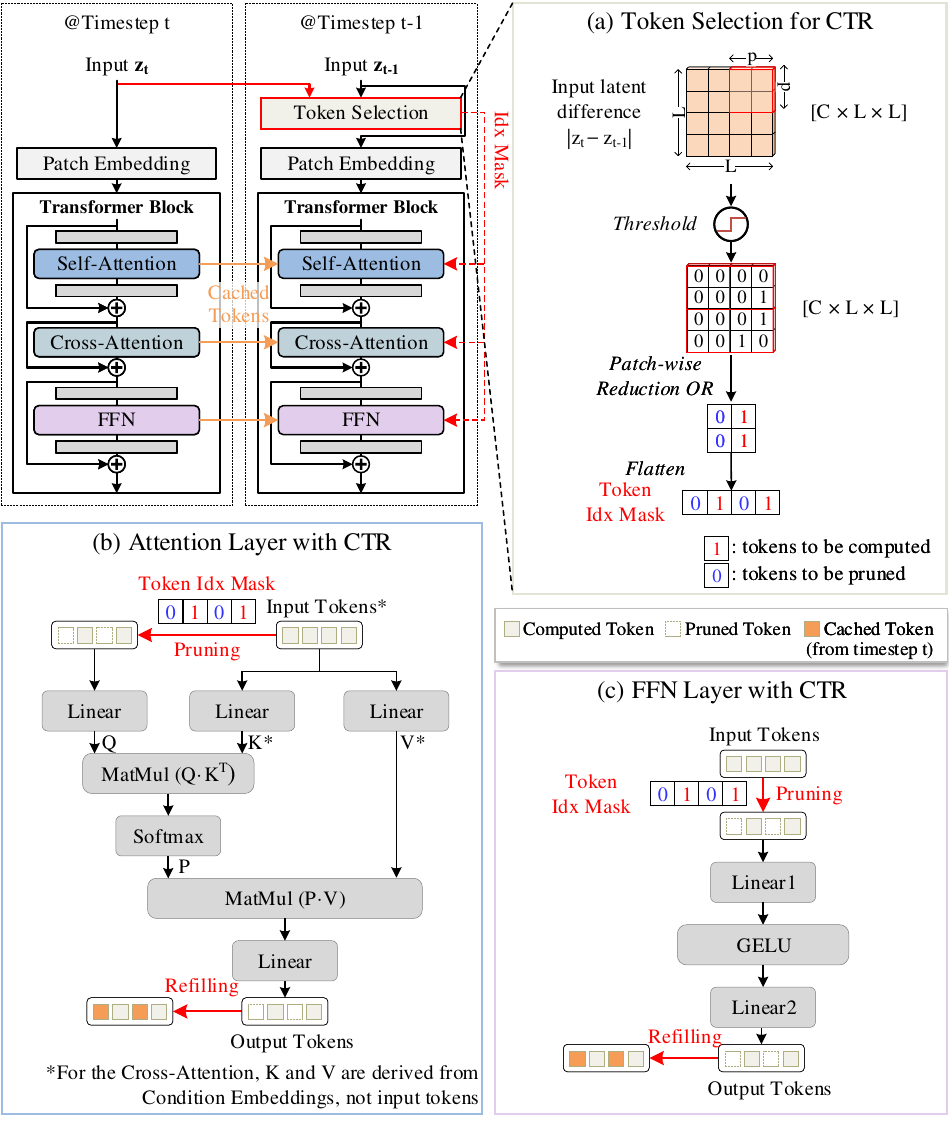}
  \caption{Cached Token Reuse (CTR) Process in timestep $t$ and $t-1$. (a) Token selection (b) Selective token computation in the attention layer (c) Selective token computation in the FFN layer}
  \label{fig9}
\end{figure}

\noindent Cached Token Reuse (CTR) is a selective computation algorithm that reuses cached token outputs in attention and FFN layers, based on changes in the input latent across timesteps. Tokens with small change reuse cached outputs from previous steps, while the remaining tokens are computed. 

The core of CTR is its token selection mechanism (Figure~\ref{fig9}(a)), which translates spatial changes in the input latent into a token-level pruning decision.
At a given timestep $t-1$, the algorithm first computes the absolute difference between the current input latent $z_{t-1}$ and the previous step's latent $z_t$.
A binary mask is then generated by applying a predefined threshold, where positions with a difference below the threshold are set to 0, and the others are set to 1.
To align this latent mask with the token structure, CTR adopts the patch-wise structure of the patch embedding process.
This binary mask is partitioned into these patches, and a reduction-OR operation is applied within each patch.
The resulting patch-wise values are flattened to produce the final token index mask, where 1 indicates tokens selected for computation and 0 indicates those to be pruned (allowing their cached values to be used).
This conservative (reduction-OR) strategy, which computes a token if any corresponding latent position shows a notable change, ensures the preservation of generation quality.

Figure~\ref{fig9}(b) and (c) illustrate how this generated token index mask is applied to the attention and FFN layers.
For the self-attention layer, the mask is used to selectively compute the query tokens: only the input tokens with a mask value of 1 are multiplied by the weight $W_Q$ to generate the corresponding query tokens.
In contrast, the key and values are computed for all tokens, regardless of the mask.
This is essential because the unpruned query tokens must still attend to the full set of key tokens (including those corresponding to pruned positions) to accurately compute their attention scores in the $QK^T$ operation.
Consequently, the computational savings begin at the query generation step. This reduction in the number of active query tokens propagates through all subsequent operations within the attention block. Specifically, the $QK^T$ matrix multiplication, the softmax computation, the $PV$ (attention-value) multiplication, and the final output projection all benefit from operating on this smaller set of query tokens.

For the cross-attention layer, the mask is similarly applied to selectively compute the query tokens derived from the input latents. The key and value are derived from text embeddings and are thus independent of the pruning mask.
For the FFN layer, the input tokens are pruned using the same token index mask, and only the non-pruned tokens are passed through the FFN computation.
After the computation in each layer (self-attention, cross-attention, and FFN), the computed outputs are integrated with the cached outputs from the previous step to reconstruct the complete set of tokens for the next operation. 
This CTR strategy effectively reduces redundant computations by 
selectively pruning tokens based on input variations.

\subsection{Softmax Thresholding with Sparsity Mask Reuse (ST)}
\label{subsec:st_algorithm}
\begin{figure}[t]
  \centering
  \includegraphics[width=\linewidth]{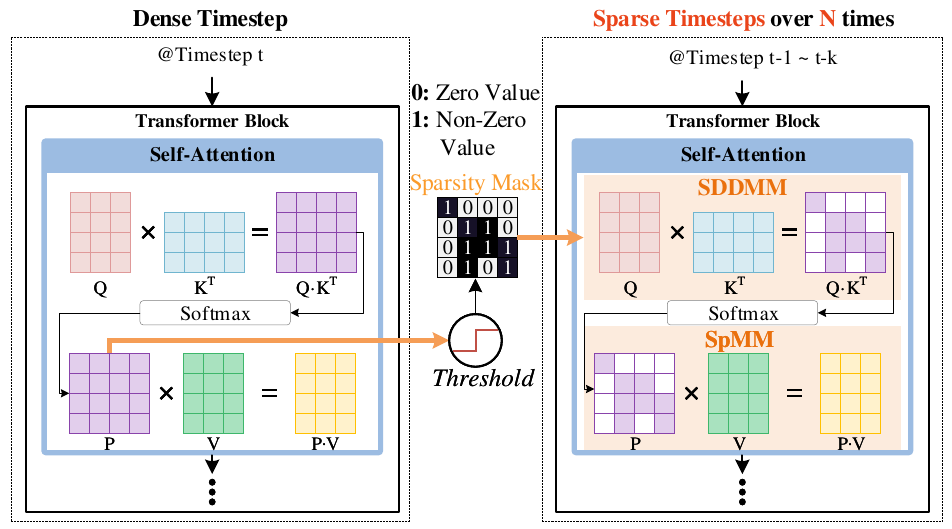}
  \caption{Process of Softmax Thresholding with Sparsity Mask Reuse}
  \label{fig10}
\end{figure}
\noindent As discussed in~\ref{subsec:st_motiv}, our approach stems from two key observations regarding the attention probability in transformer-based diffusion models: a significant portion of the values are negligible, and their sparsity patterns remain highly similar across adjacent timesteps.
These observations motivate our Softmax Thresholding with Sparsity Mask Reuse (ST) strategy. 
The prevalence of near-zero values allows us to apply a threshold to induce sparsity, while the temporal similarity of these patterns enables the reuse of this sparsity mask over several subsequent timesteps.
The ST process, illustrated in Figure~\ref{fig10}, operates by alternating between dense step and sparse steps.
In a dense step, the sparsity mask is generated.
The attention probability matrix $P$ is computed and then apply a binary thresholding operation using a pre-determined threshold $\tau$ to create a binary sparsity mask $M$:
\[
\text{Sparsity Mask} =
\begin{cases}
1, & \text{if } S_{ij} \geq \tau \\
0, & \text{otherwise}
\end{cases}
\]
In the subsequent sparse steps, this mask $M$ is reused to optimize two operations.
First, the dense $QK^T$ matrix multiplication is replaced with a Sampled Dense-Dense Matrix Multiplication (SDDMM).
Using $M$ as a guide, this operation only computes the dot products corresponding to the non-zero entries in the mask.
Second, after the sparse $QK^T$ and softmax, the resulting sparse attention probability matrix $P_{\text{sparse}}$ is multiplied with the Value matrix $V$.
This step ($P_{\text{sparse}} \cdot V$) becomes a Sparse-Dense Matrix Multiplication (SpMM), which requires significantly fewer computations than the original dense-dense operation.
Together, these transformations from dense to sparse formulations (SDDMM + SpMM) significantly reduce redundant operations in the attention layer while minimizing accuracy loss.
\section{Specialized Hardware Architecture}
\subsection{Architecture Overview}
\begin{figure}[t]
  \centering
  \includegraphics[width=\linewidth]{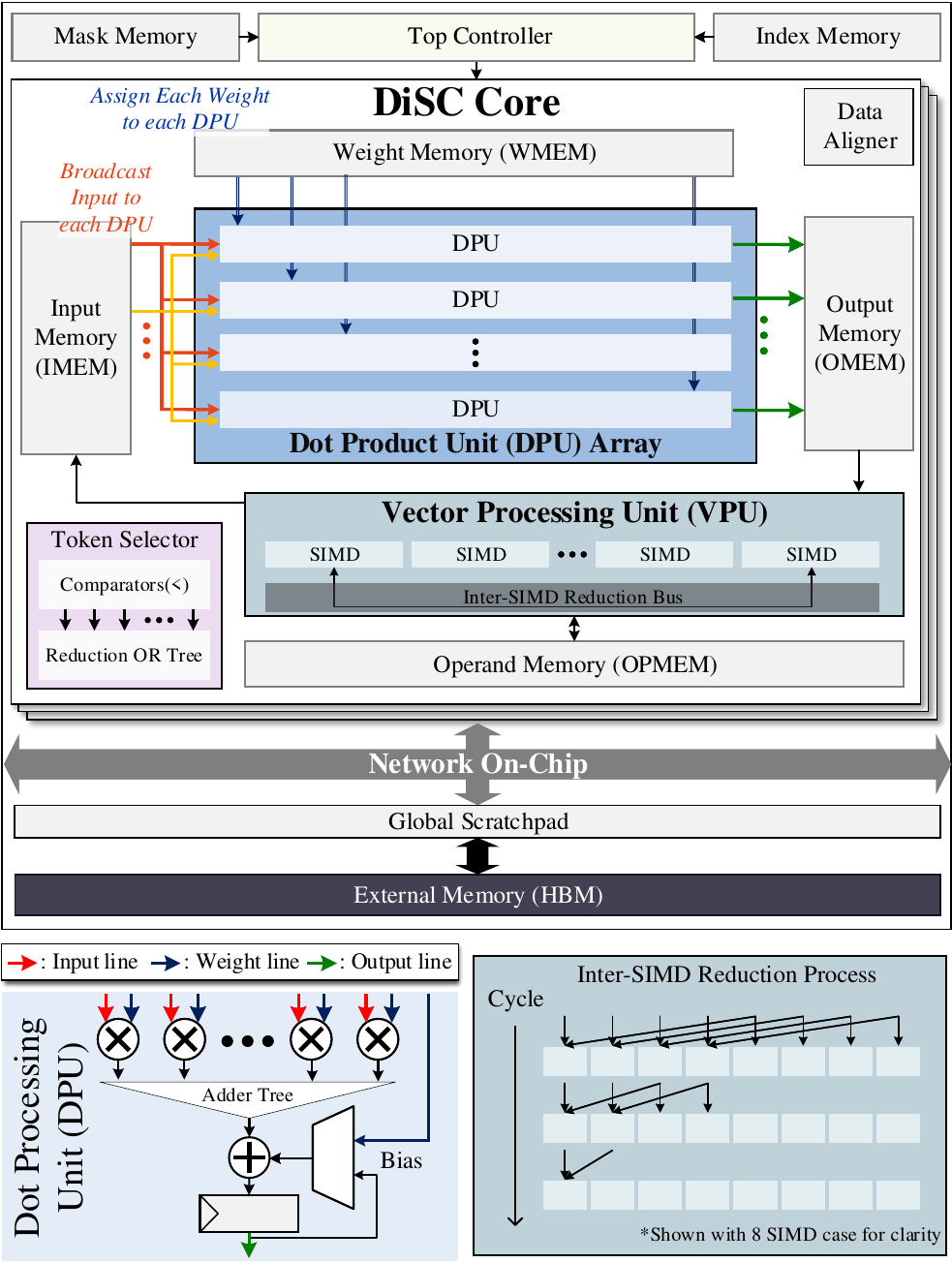}
  \caption{Overview of DiSC Architecture}
  \label{fig12}
\end{figure}

\noindent Figure~\ref{fig12} illustrates the overall hardware architecture of DiSC, which is co-designed to execute CTR and ST.
Each DiSC core comprises a Dot Product Unit (DPU) array for matrix multiplications, a Vector Processing Unit (VPU) for non-matrix operations, and specialized support unit including a Token Selector and Data Aligner.
A top controller orchestrates the core's operations, using sparsity masks and token indices (from Mask/Index Memory) to generate control signals.
Outside the cores, a global scratchpad (GSC) acts as an intermediate data buffer between the cores and external HBM, connected via a Network-on-Chip (NoC).

\textbf{Dot Product Unit Array:} 
The DPU array is designed to perform matrix multiplications. 
It is is composed of a set of DPUs, and each DPU contains multipliers, an adder tree, and an accumulation register, fetching inputs and weights from on-chip memories (Input/Weight Memory) and storing results in Output Memory. 

\textbf{Vector Processing Unit:}
The VPU handles all vector and element-wise operations (e.g., softmax, GELU, LayerNorm) using multiple SIMD engines, each paired with an on-chip SRAM bank. 
It integrates an inter-SIMD reduction bus to support cross-bank accumulations, a feature essential for operations like the softmax denominator or LayerNorm's mean/variance.
As we will show, this existing bus is opportunistically reused for SpMM accumulation.

\textbf{Token Selector:}
The Token Selector is a lightweight hardware unit that implements the CTR token selection logic.
After the VPU computes the latent differences, the Token Selector applies the threshold $\tau$ via comparators and executes the patch-wise OR reduction using a reduction-OR tree to generate the final token index mask.

\subsection{Sparse Operation Flow}
DiSC's datapath is designed to handle CTR and ST's distinct patterns with minimal overhead.
For CTR, which prunes at the token-level, operations are efficiently skipped without any hardware modifications.
Both input memory and operand memory are mapped row-major, with the row dimension corresponding to tokens.
The controller simply consults the token index mask and issues on-chip memory addresses only for non-pruned tokens, effectively gating computation for pruned tokens without cost.
\begin{figure}[t]
  \centering
  \includegraphics[width=\linewidth]{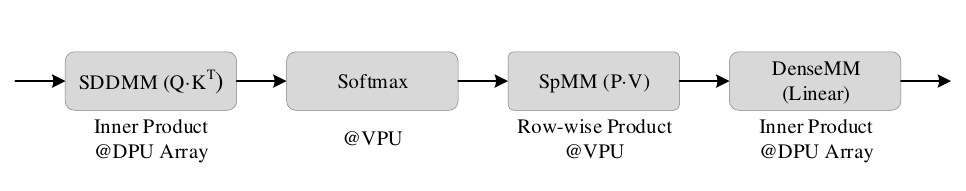}
  \caption{Sparse Operation Flow}
  \label{fig14}
\end{figure}
For ST, which induces element-level sparsity, the flow is handled differently, as shown in Figure~\ref{fig14}.
ST transforms dense attention operations into a sequence of sparse operations: (1) $QK^{\mathsf{T}}$ becomes a Sampled Dense-Dense Matrix Multiplication (SDDMM), and (2) $P \cdot V$ becomes a Sparse-Dense Matrix Multiplication (SpMM).

A key aspect of DiSC's design is how it executes this sequence efficiently without dedicated sparse processing units.
The SDDMM operation is executed on the DPU array.
While this operation utilizes the independent access capability of the on-chip multi-bank SRAMs, a hash-based distribution (Section~\ref{subsec:hash_flow}) is employed to address the load-balancing problem across the DPUs.

The subsequent SpMM operation presents a critical design choice. 
Prior work on sparse accelerators~\cite{srivastava2020matraptor, zhang2021gamma} often employs dedicated units for sparsity handling, which are optimized for highly sparse matrices.
However, this approach is suboptimal for diffusion models, which present a hybrid workload of dense and moderately sparse operations; a dedicated sparse unit would be under-utilized during dense computations.
Therefore, DiSC adopts a more area-efficient strategy: the SpMM operation is offloaded to the VPU.
This solution repurposes the VPU, already required for vector operations, to efficiently handle the moderate-sparsity SpMM using its existing row-wise product dataflow.

This flow also provides a critical pipelining benefit.
In the original dense flow, consecutive matrix multiplications ($P \cdot V$ and the output projection) would both execute sequentially on the DPU array, causing stalls.
In DiSC's flow, the VPU execute the SpMM operation ($P \cdot V$), and the DPU can begin the next linear operation (which consumes the SpMM output) immediately, reading results from the on-chip memory as they are produced.
This overlaps the execution of the two units, effectively hiding the latency of the second operation and improving overall hardware utilization.

\subsection{Executing Sparse Operations with Hash-based Distribution}
\label{subsec:hash_flow}
\begin{figure}[t]
  \centering
  \includegraphics[width=\linewidth]{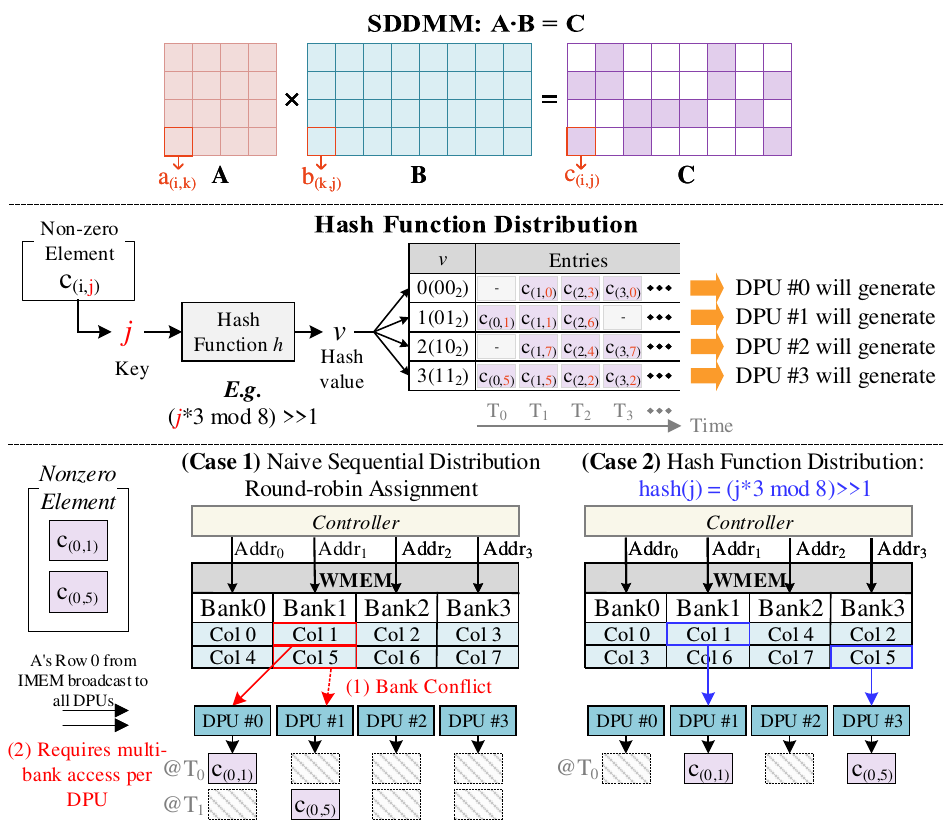}
  \caption{SDDMM execution with Hash-based Distribution}
  \label{fig16}
\end{figure}
\begin{figure}[t]
  \centering
  \includegraphics[width=\linewidth]{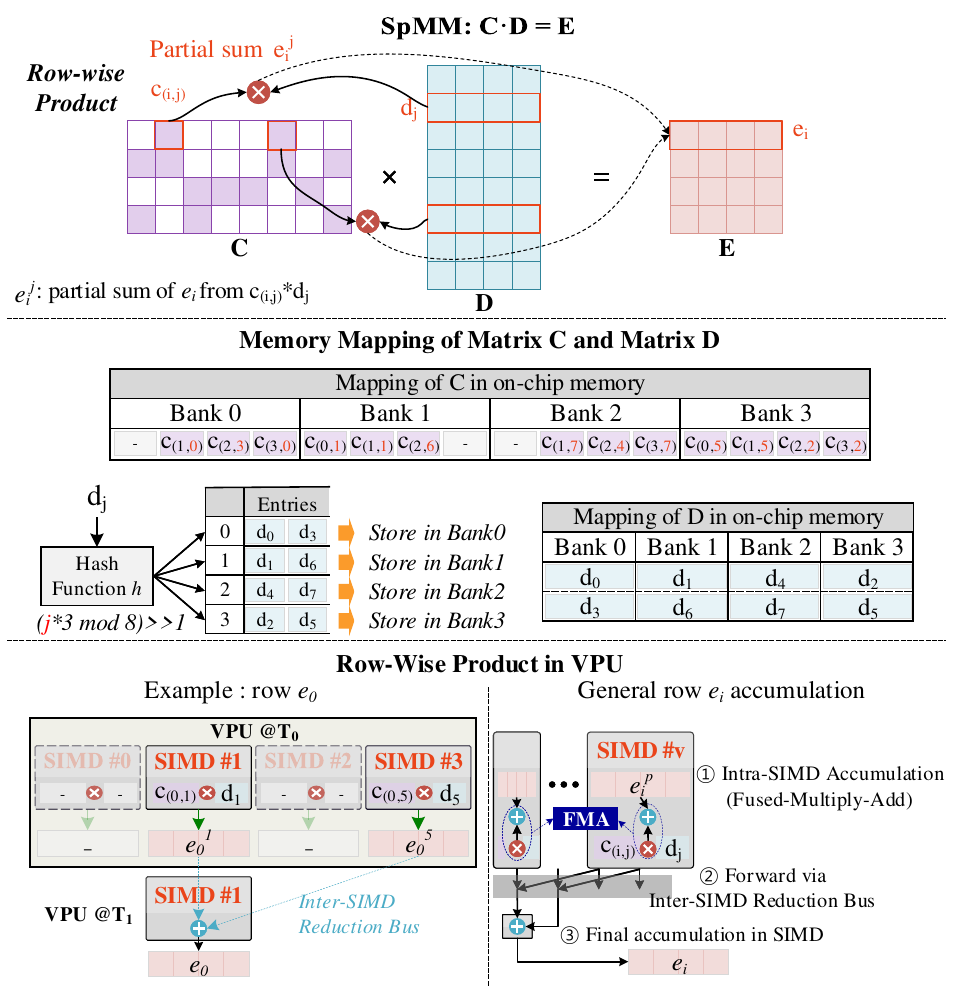}
  \caption{SpMM Execution with Hash-based Distribution}
  \label{fig17}
\end{figure}
A critical challenge in executing sparse operations is to handle both SDDMM and SpMM efficiently without dedicated sparse hardware.
DiSC solves this by implementing a unified dataflow where both operations share a common hash-based distribution, as shown in Figures~\ref{fig16} and~\ref{fig17}.
This strategy's goal is to solve the critical load-balancing problem of sparse operations while avoiding costly sparse format decoders and interconnection networks.

First, the SDDMM operation ($QK^{\mathsf{T}}$) is executed on the DPU array.
The primary challenge is to map sparse computations to parallel DPUs without incurring bank conflicts or requiring a complex interconnection network.
Figure~\ref{fig16} illustrates how our hash-based distribution strategy solves both problems.
A naive round-robin approach (Case 1), which might sequentially assign non-zeros to DPUs (e.g., 1st non-zero to DPU 0, 2nd to DPU 1, ...), can cause bank conflicts if different DPUs require columns (e.g., col 1 and col 5) that are stored in the same memory bank.
It also necessitates a costly interconnection network so that any DPU can access any bank.
Our strategy (Case 2) avoids this by defining two rules based on a hash function ($v = hash(j)$):
\begin{itemize}
\item \textbf{Weight Mapping:} The dense weight column $j$ (from $K^{\mathsf{T}}$) is pre-distributed into WMEM Bank $v$.
\item \textbf{Computation Assignment:} The non-zero output $c_{(i,j)}$ is computed exclusively by DPU $v$.
\end{itemize}

This mapping simultaneously solves both primary challenges: it (1) eliminates the need for an all-to-all interconnection network, as DPU $v$ is only required to access its dedicated WMEM Bank $v$, and (2) inherently prevents bank conflicts, as no two different DPUs can access the same memory bank.
This parallel operation is enabled by the independent access capability of the on-chip SRAM banks, allowing each DPU-bank pair (e.g., DPU1-Bank1, DPU3-Bank3) to fetch data from different addresses simultaneously.
This strategy, combined with a well-chosen hash function, also addresses the separate problem of load balancing by distributing the number of non-zero elements evenly across DPUs, which we analyze in Section~\ref{subsec:additional_details}.

The output of SDDMM is stored in OMEM in a hash-encoded format, where an element's bank location implicitly encodes its hash value.
This encoded representation is the key to the unified flow.
After passing through the softmax function (which we detail in Section~\ref{subsec:additional_details}), this hash-encoded matrix $C$ is directly consumed by the subsequent SpMM operation.

This SpMM operation ($P \cdot V$) is offloaded to the VPU, as illustrated in Figure~\ref{fig17}.
This avoids any format conversion or decoding overhead, because the VPU's dataflow is designed to natively operate on the hash-encoded data.
The solution relies on the same hash function used by the DPU:
\begin{itemize}
\item \textbf{Sparse Input:} The hash-encoded matrix $C$ is already in the on-chip memory, with bank $v$ holding non-zero elements $c_{(i,j)}$ where $hash(j) = v$
\item \textbf{Dense Input:} The dense matrix $D$ (the Value matrix $V$) is also mapped such that bank $v$ stores all rows $d_j$ where $hash(j) = v$.
\end{itemize}

This consistent mapping allows the VPU's SIMD $v$ to fetch all its required operands (i.e., $c_{(i,j)}$ and $d_j$) by exclusively accessing its corresponding memory bank $v$.
As with the DPU, this eliminates the need for a costly interconnection network between SIMDs and memory banks.
The VPU executes a row-wise product dataflow.
The core operation is the computation of partial sums $e_i^j$ (where $e_i^j = c_{(i,j)} \times d_j$).
This operation maps perfectly to the Fused-Multiply-Add (FMA) capability of the SIMD lanes, which computes this product and adds it to an accumulator in a single operation, allowing the VPU to operate at its peak computational throughput.
To form the final output row $e_i$, accumulation occurs in two stages:
First, partial sums are accumulated within each SIMD using this FMA mode. 
Second, the resulting accumulated values from each SIMD are aggregated using the VPU's existing inter-SIMD reduction bus.

In summary, this unified hash-based distribution enables a seamless, zero-decode pipeline. 
It allows the DPU and VPU to handle SDDMM and SpMM, respectively, while avoiding dedicated hardware for sparse format decoding, interconnection, or reduction.
\subsection{Additional Implementation Details}
\label{subsec:additional_details}
\begin{figure}[t]
  \centering
  \includegraphics[width=\linewidth]{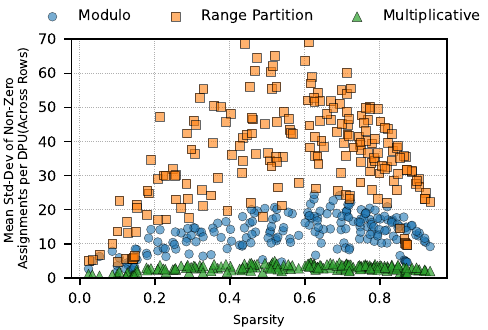}
  \caption{Load Balancing Across DPUs for Different Hash Functions. The results are obtained using the sparsity patterns from the PixArt‑$\Sigma$ model at 2k$\times$2k resolution after applying ST. We use N=64 and set the $k_{odd}$= 2053.}
  \label{fig_load}
\end{figure}
\textbf{Hash function.} To analyze the impact of different hash functions $hash(j)$ on workload distribution, we verified with three candidate functions for mapping a column index j to a DPU, where $N$ denotes the number of DPUs:
\begin{itemize}
    \item \textbf{Modular hash:} defined as $j~mod~N$ 
    \item \textbf{Multiplicative hash:} defined as $MSB_b(j*k_{odd})$, where $k_{odd}$ is an odd constant, $MSB_b(x)$ extracts the $b$ most significant bits of $x$ (with $b = \log_2 N$)
    \item \textbf{Range partition:} which divides each row into N contiguous segments, equivalent to $MSB_b(j)$ where $b = \log_2 N$
\end{itemize}
For this analysis, we obtained sparsity masks by applying ST to the PixArt-$\Sigma$ model at 2k$\times$2k resolution.
For each block and diffusion timesteps, we applied the hash function to every row of the sparsity mask and counted, for each DPU, the number of non-zero elements assigned to it.
We then computed the standard deviation of these counts across DPUs for each row and averaged the results over all rows.
Figure~\ref{fig_load} plots the average standard deviation of the number of non-zero elements assigned to each DPU as a function of matrix sparsity for the three hash functions.
As shown in the figure, the three hash functions show similar standard deviations when sparsity is either very high or very low.
However, in the moderate‑sparsity range, the multiplicative hash consistently yields smaller standard deviations, indicating a more balanced distribution. Therefore, we adopt the multiplicative hash function for subsequent performance evaluations.

\textbf{Softmax.} 
In between the SDDMM and SpMM operations, the softmax function is applied to the hash‑encoded matrix $C$. To compute the denominator of the softmax (i.e., the sum of exponentials) for a given row, the controller asserts valid signals only for the elements in that row. Because the VPU already incorporates both intra‑SIMD accumulation paths and an inter‑SIMD reduction bus, the required summation can be performed without any additional hardware support.

\begin{figure}[t]
  \centering
  \includegraphics[width=\linewidth]{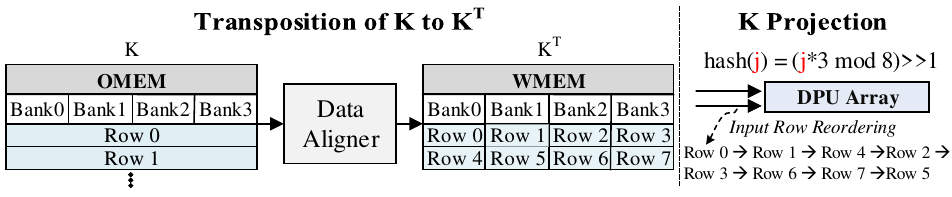}
  \caption{Data Alignment and Row Reordering for Hash‑Based Mapping.}
  \label{fig_align}
\end{figure}
\textbf{Memory Mapping for Hash-based Distribution.}
Applying the hash-based distribution requires the on-chip memory layout of matrix $B$ (SDDMM, $K^T$) and matrix $D$ (SpMM, $V$) to differ from the dense layout.
DiSC achieves this complex mapping efficiently by modifying the preceding $K$ and $V$ projection step, as illustrated in Figure~\ref{fig_align}.
During the $K$ projection, the controller modifies the order of computation, generating the rows of $K$ in a pre-shuffled sequence (e.g., Row 0, 1, 4, 2, 3...) and storing them in OMEM.
This sequence is precisely determined so that when the existing Data Aligner performs its transposition operation ($K \rightarrow K^T$), the $K^T$ columns naturally land in the correct hash-mapped WMEM banks (e.g., Bank 0 holding Row 0 and Row 3).
A similar controller-driven pre-ordering is applied during the $V$ projection, ensuring the rows of $V$ are also correctly mapped to their corresponding banks for the SpMM operation.
This approach is highly efficient as it achieves the required layout, relying only on modified controller addressing and the existing transposition function of the Data Aligner.
\section{Evaluation}
\subsection{Methodology}
\begin{table}[t]
  \centering
  \includegraphics[width=\linewidth]{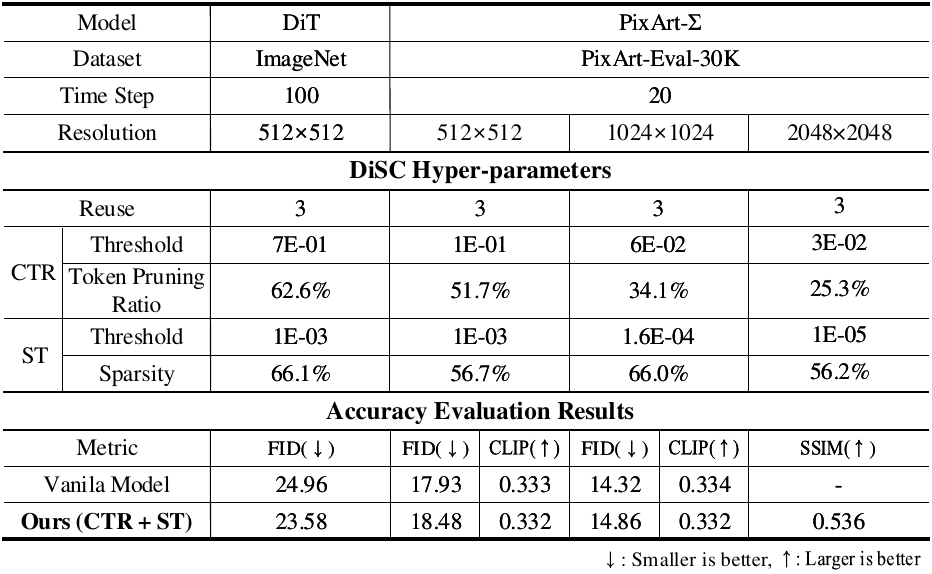}
  \caption{Model Setup and Accuracy Evaluation Results}
  \label{table1}
\end{table}

\noindent \textbf{Workloads.}
To evaluate the accuracy and performance of DiSC across various image resolutions, we utilize two diffusion models.
Specifically, the DiT~\cite{peebles2023scalable} is evaluated at 512$\times$512 image resolution, and the PixArt-$\Sigma$~\cite{chen2024pixart} is evaluated at 512$\times$512, 1k$\times$1k, and 2k$\times$2k image resolutions.
Since our objective is to accelerate high-resolution image generation, we exclude the 256$\times$256 image resolution from our evaluation.
The datasets used are ImageNet~\cite{krizhevsky2012imagenet} for DiT and PixArt-Eval-30K~\cite{pixart_eval_30k} for PixArt-$\Sigma$.
Each model is evaluated using its default inference settings as provided in the official implementation.

\textbf{Accuracy.}
DiSC proposes the CTR and ST algorithms, which reduce computations to accelerate the diffusion model's denoising process.
A critical part of our evaluation is to validate that these optimization strategies preserves the model's accuracy.
We use the Frechet Inception Distance (FID)~\cite{heusel2017gans} to measure image fidelity, and the CLIP Score~\cite{hessel2021clipscore} to evaluate semantic alignment. 
For the 2k$\times$2k resolution case, due to the prohibitive GPU runtime when applying both CTR and ST, we instead sample 50 outputs and report the Structural Similarity Index (SSIM) as an additional reference metric for perceptual quality.
The hyperparameters for CTR and ST were tuned on a randomly sampled 10\% subset of the dataset. We selected the values that provided the best trade-off between generation quality (e.g., FID) and the target computational reduction (i.e., token pruning ratio and sparsity).
The specific hyperparameters selected for each configuration are detailed in Table~\ref{table1}.

\textbf{Hardware.}
To evaluate the performance of the DiSC hardware, we implemented a custom cycle-level simulator that leverages the event-driven simulation framework of gem5~\cite{binkert2011gem5} and integrates with Ramulator 2.0~\cite{Luo2024ramulator2}.
We use the NVIDIA A100 80~GB PCIe~\cite{nvidiaA100} as a comparison baseline and measure latency using the NVIDIA Nsight Systems.
For a fair comparison, we scale out the DiSC hardware to match the peak performance and GPU memory bandwidth of the A100.
The A100's FP16 Tensor Cores offer a peak MAC throughput of 312~TFLOPS. To match this, we deploy 38 DiSC cores, each providing a peak MAC throughput of 8.2~TFLOPS, resulting in a total of 311.3~TFLOPS of peak MAC performance.
Regarding memory bandwidth, the A100 has a bandwidth of 1935 GB/s, so we configure eight stacks of HBM2, each with eight channels, achieving a total bandwidth of 2048 GB/s. 

\textbf{Area and power measurement.}
To measure the area and power of DiSC, we implement the design in SystemVerilog and synthesize it at 1 GHz using Synopsys Design Compiler~\cite{synopsysDC} and Samsung’s 28nm standard cell library. The on-chip SRAM memories were compiled using the same 28nm library. GPU energy consumption is measured using NVIDIA-SMI, while external memory and on-chip memory energy are estimated using~\cite{OConnor2017FineGrainedDRAM} and CACTI~\cite{balasubramonian2017cacti}.

\subsection{Evaluation Results}
\textbf{Accuracy.}
Table~\ref{table1} presents the hyperparameter configurations and accuracy results.
As shown, DiSC achieves substantial computational savings---token pruning ratio from 25.3\% to 62.6\% and sparsity from 56.2\% to 66.1\%--- with only a negligible impact on generation quality.
In terms of image quality, all models and resolutions exhibit a negligible increase in FID (less than 0.6) compared to the vanilla baseline.
CLIP Score also shows no meaningful degradation.
These results confirm that CTR and ST introduce only a minimal impact on accuracy, validating our optimization strategies.
\begin{figure}[t]
  \centering
  \includegraphics[width=\linewidth]{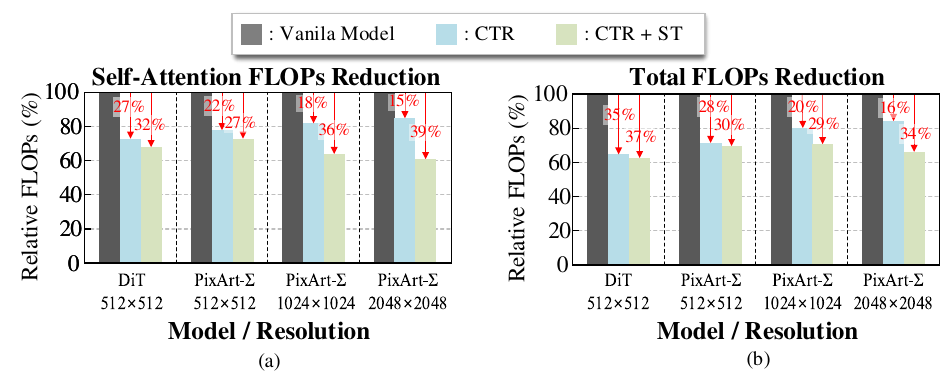}
  \caption{FLOPs Reduction in (a) Self-Attention layer and (b) total steps of Denoising Network}
  \label{fig18}
\end{figure}

\textbf{Reduction in FLOPs.}
Figure~\ref{fig18} illustrates the FLOPs reduction from CTR and ST, showing their contributions in self-attention (a) and the entire network (b).
The two schemes show complementary benefits based on resolution.
CTR (applied to FFN, cross-attention, and self-attention) provides a larger reduction at lower resolutions, where FFN layers are dominant relative to self-attention.
As resolution increases, self-attention ($QK^T, PV$) becomes the primary bottleneck due to its quadratic ($O(N^2)$) computational complexity, and the benefits of ST become critically important.
While the relative token pruning ratio of CTR decreases at higher resolutions, ST's effectiveness is amplified.
This effect is significant: in PixArt-$\Sigma$ (Fig~\ref{fig18}(a)), the FLOPs reduction attributable to ST jumps from 5\% (at 512) to 24\% (at 2k).
Overall (Fig~\ref{fig18}(b)), the combination achieves a 29-37\% total FLOPs reduction.
These results validate that supporting both CTR and ST is essential for achieving substantial acceleration, especially as self-attention dominates computation at high resolutions.

\begin{figure}[t]
  \centering
  \includegraphics[width=\linewidth]{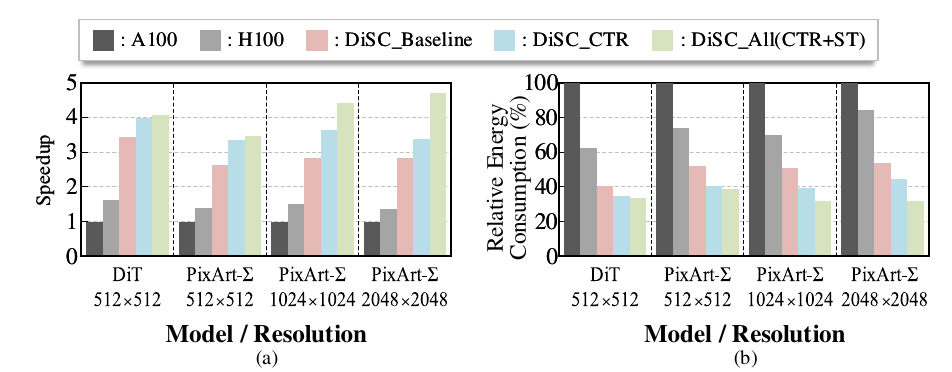}
  \caption{(a) Speedup over NVIDIA A100 GPU (b) Relative Energy Consumption over NVIDIA A100}
  \label{fig19}
\end{figure}

\textbf{Speedup.}
Figure~\ref{fig19}(a) presents the speedup relative to the NVIDIA A100 GPU baseline.
We conducted an ablation study using three configurations: \text{DiSC\_Baseline}, \text{DiSC\_CTR}, and \text{DiSC\_All}.
\text{DiSC\_Baseline} is the configuration that operates on the DiSC hardware without applying either CTR or ST.
\text{DiSC\_CTR} applies only CTR, and \text{DiSC\_All} combines both CTR and ST.
First, the specialized DiSC hardware alone (\text{DiSC\_Baseline}) provides a substantial 2.64$-$3.44$\times$ speedup over the A100, demonstrating the efficiency of our custom datapath.
Applying CTR (\text{DiSC\_CTR}) further boosts this performance to 3.41$-$4.01$\times$.
Finally, \text{DiSC\_All} achieves the highest speedup of 3.47$-$4.74$\times$. 
Notably, \text{DiSC\_All} also outperforms NVIDIA’s H100 GPU, achieving a 2.48$-$3.50$\times$ speedup.
This confirms that our hardware-software co-design effectively translates FLOPs reduction into substantial latency improvements.

\textbf{Energy Consumption.}
Figure~\ref{fig19}(b) shows the energy consumption relative to the A100 baseline.
The specialized \text{DiSC\_Baseline} hardware achieves an energy reduction of 46.4$-$59.3\%.
Applying CTR (\text{DiSC\_CTR}) increases this reduction to 55.3$-$65.4\%, and \text{DiSC\_ALL}, which combines both CTR and ST, achieves a total reduction of 61.3-68.1\%.
This energy efficiency advantage extends even to the current-generation NVIDIA H100 GPU, as \text{DiSC\_All} consumes 46.4$\text{--}$62.1\% less energy.
These results demonstrate the effectiveness of the proposed DiSC hardware in reducing energy consumption. 
In particular, when the sparsity induced by ST leads to SpMM, executing such operations on the DPU can result in low utilization.
By offloading these operations to the VPU, the utilization is improved, which in turn leads to a significant reduction in energy consumption.
This indicates that the DiSC architecture is well optimized for the computation pattern changes introduced by CTR and ST.
\subsection{Area and Power Breakdown}

\begin{table}[t]
  \centering
  \includegraphics[width=\linewidth]{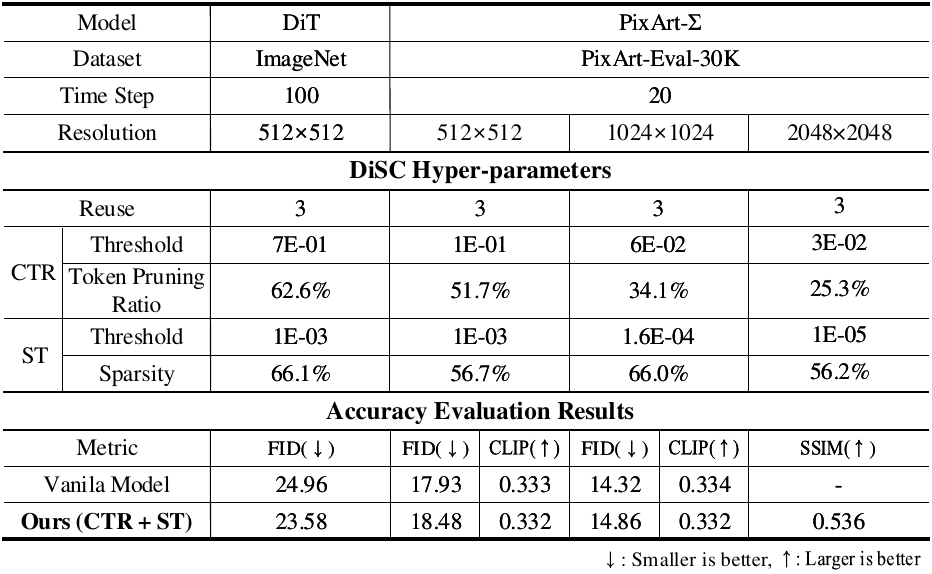}
  \caption{DiSC Area and Power Breakdown in Samsung 28nm}
  \vspace{-0.3in}
  \label{table2}
\end{table}

Table~\ref{table2} presents the area and power breakdown of a DiSC core.
A key architectural advantage of DiSC is its area-efficient method for exploiting sparsity.
This efficiency is achieved through a hash-based distribution of non-zero elements across multi-bank SRAMs that support independent access.
This mechanism effectively solves the critical load-balancing problem inherent in sparse operations, thereby mitigating the risk of under-utilization without requiring dedicated sparse processing units.
This demonstrates a core principle of our co-design: exploiting complex unstructured sparsity patterns with minimal hardware modifications.
For a chip-level comparison, the NVIDIA A100 (fabricated in a 7nm process) occupies a total die area of 826mm².
In contrast, our full DiSC design—including 38 cores and 40MB of global scratchpad memory—has a total area of 596.8mm² in a 28nm process.
To enable a fair comparison across these different technology nodes, the DiSC design's area is scaled to a 7nm process using technology scaling principles~\cite{Sarangi2021DeepScaleTool}, resulting in an estimated area of 70.86mm².
This area gap stems from the area efficiency of DiSC's specialized architecture, which omits the general-purpose overhead of the A100 that is unnecessary for diffusion models.

\subsection{Comparison with Prior Diffusion Schemes at High Resolution}
\label{sec5.5}
\begin{figure}[t]
  \centering
  \includegraphics[width=\linewidth]{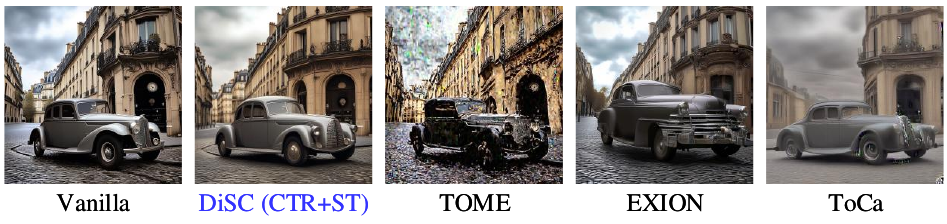}
  \caption{Visual comparison of generated images on PixArt-$\Sigma$ at 1k$\times$1k resolution.}
  \label{fig_existing}
\end{figure}
\begin{table}[t]
  \centering
  \includegraphics[width=\linewidth]{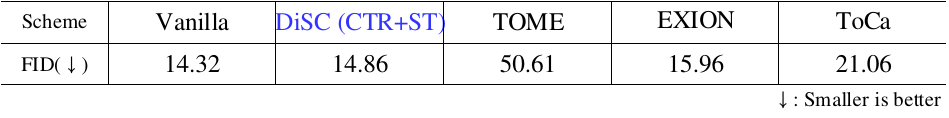}
  \caption{FID score comparison with prior schemes on PixArt-$\Sigma$ at 1k$\times$1k resolution. }
  \vspace{-0.2in}
  \label{Table3}
\end{table}

Several prior works have proposed techniques to reduce the computational cost of attention for diffusion models, such as Token Merging (TOME)~\cite{bolya2023token}, EXION~\cite{heo2025exionexploitinginterintraiteration}, and ToCa~\cite{zou2024accelerating}.
However, these schems were typically demonstrated on U-Net-based diffusion models (TOME) or at lower resolutions (256$\times$256 for EXION and ToCa).

As self-attention becomes the dominant computational bottleneck at high resolutions, our objective was to evaluate the scalability of their self-attention optimization strategies on the demanding 1k$\times$1k PixArt-$\Sigma$ workload.
Therefore, for the prior works, we applied their respective self-attention schemes.
For TOME, we applied token merging only to the self-attention layers with a 10\% token reduction ratio ($r=10\%$).
For EXION, we applied its eager prediction scheme, which involves approximating attention scores using its leading-one detection (LOD) method.
We then applied top-k sparsity to these approximated scores (i.e., keeping only the top 5\% per row).
We used $k=0.05$, a hyperparameter adopted from the original paper's DiT experiments, and omitted an additional threshold-based approximation, which requires workload-specific tuning.
For ToCa, we implemented its self-attention strategy (caching the entire layer output with $N=3$) and omitted its token-level FFN/cross-attention caching, which requires workload-specific hyperparameter tuning.

The results are summarized in Figure~\ref{fig_existing} and Table~\ref{Table3}.
As shown in Figure~\ref{fig_existing}, prior schemes exhibit visible artifacts when scaled to 1k$\times$1k resolution.
The FID scores, measured on 5k samples, quantify this degradation.
Compared to the Vanilla baseline (14.32), TOME's FID increases to 50.61, while EXION (15.96) and ToCa (21.06) also show accuracy degradation.
In contrast, DiSC (14.86) achieves the smallest FID degradation (+0.54) among the evaluated schemes, demonstrating its ability to preserve image quality at high resolution.

\section{Related Work}
\subsection{Acceleration for Diffusion Models}
\label{subsec:acc}
\noindent One of the representative approaches for accelerating diffusion models is caching-based reuse.
Early studies such as FORA~\cite{selvaraju2024fora}, DeepCache~\cite{ma2024deepcache}, and Cache Me If You Can~\cite{wimbauer2024cache} exploited the temporal similarity in diffusion processes by reusing layer-level feature maps.
More recently, token-level caching schemes such as ToCa~\cite{zou2024accelerating} have emerged that selectively cache important tokens based on importance estimation.
However, ToCa's token-level caching is limited to FFN and cross-attention layers; for self-attention, it reuses the entire layer output rather than applying token-level pruning.
In contrast, CTR introduces a mechanism that translates spatial variations in the input latent into a fine-grained, token-level pruning decision, enabling effective token-level pruning across all 3 layers, including self-attention.
Regarding resolution, most prior works~\cite{selvaraju2024fora, zou2024accelerating, zou2024DuCa} focus only on low resolutions (256$\times$256).
Although Learning-to-Cache~\cite{ma2024learning} and DiTFastAttn~\cite{yuan2024ditfastattn} report results beyond 512$\times$512 resolutions, the former requires additional hyperparameter training, and the latter incurs extra cost to determine the compression plan.
In contrast, DiSC achieves resolution-scalable by jointly applying CTR and ST.

Sparse VideoGen~\cite{xi2025sparsevideogenacceleratingvideo} accelerates video diffusion models by classifying attention heads into spatially sparse and temporally sparse. 
However, the sparsity it exploits is static and block sparsity—unlike the dynamic, unstructured sparsity in DiSC—making it compatible with GPU block-sparse libraries~\cite{guo2024blocksparse}.
This approach relies on the frame-level structure of video and is thus inapplicable to image diffusion models.

On the hardware side, prior works such as EXION~\cite{heo2025exionexploitinginterintraiteration}, Cambricon-D~\cite{kong2024cambricon}, and Ditto~\cite{kim2025ditto} have explored for diffusion models.
Among them, Cambricon-D and Ditto leverage temporal similarity across diffusion steps by quantizing the inter-step differences and using them to compute the next-step outputs.
However, applying differential computation is challenging for non-linear operations such as attention.
Moreover, the need to access previous-step activations when computing differences incurs additional memory overhead, which becomes increasingly significant as the image resolution grows.

\subsection{Hardware acceleration for sparse matrix-multiplication.}
Several prior works~\cite{pal2018outerspace, zhang2020sparch, srivastava2020matraptor, zhang2021gamma} have focused on accelerating General Sparse Matrix-Matrix Multiplication (SpGEMM), typically targeting workloads with extremely high sparsity.
However, these approaches are not well suited for the sparsity workload induced in DiSC, which exhibits moderate and hybrid dense-sparse workload.

Other prior works like Sanger~\cite{lu2021sanger} and Sigma~\cite{qin2020sigma}, support both sparse and dense operations.
However, they handle sparsity by enabling multiple PE to access the same data through multiplexing or introducing additional on-chip interconnect networks among PEs. Such additional structure introduce non-trivial hardware overhead and are less suitable for transformer-based diffusion models, which only exhibit moderate sparsity.
\section{Conclusion}
\noindent In summary, we propose DiSC, a resolution-scalable, sparsity-aware HW-SW co-designed accelerator targeting transformer-based diffusion models.
Our software algorithms, Cached Token Reuse (CTR) and Softmax Thresholding with Sparsity Mask Reuse (ST), jointly exploit temporal similarity to induce a moderate sparsity and hybrid dense-sparse workload.

To efficiently handle this specific workload, our specialized hardware architecture avoids dedicated sparse units, instead repurposing its existing compute engines via a hash-based distribution.
As a result, DiSC achieves up to 3.47-4.74$\times$ speedup and 61.3-68.1\% energy saving over the NVIDIA A100 GPU, and 2.48–3.5$\times$ speedup and 46.4–62.1\% energy saving over the H100.

\bibliographystyle{IEEEtran}
\bibliography{refs}
{\raggedbottom
\begin{minipage}[t]{\columnwidth}
\vspace*{-20pt}
\begin{IEEEbiography}
[{\includegraphics[width=1in,height=1.25in,clip,keepaspectratio]{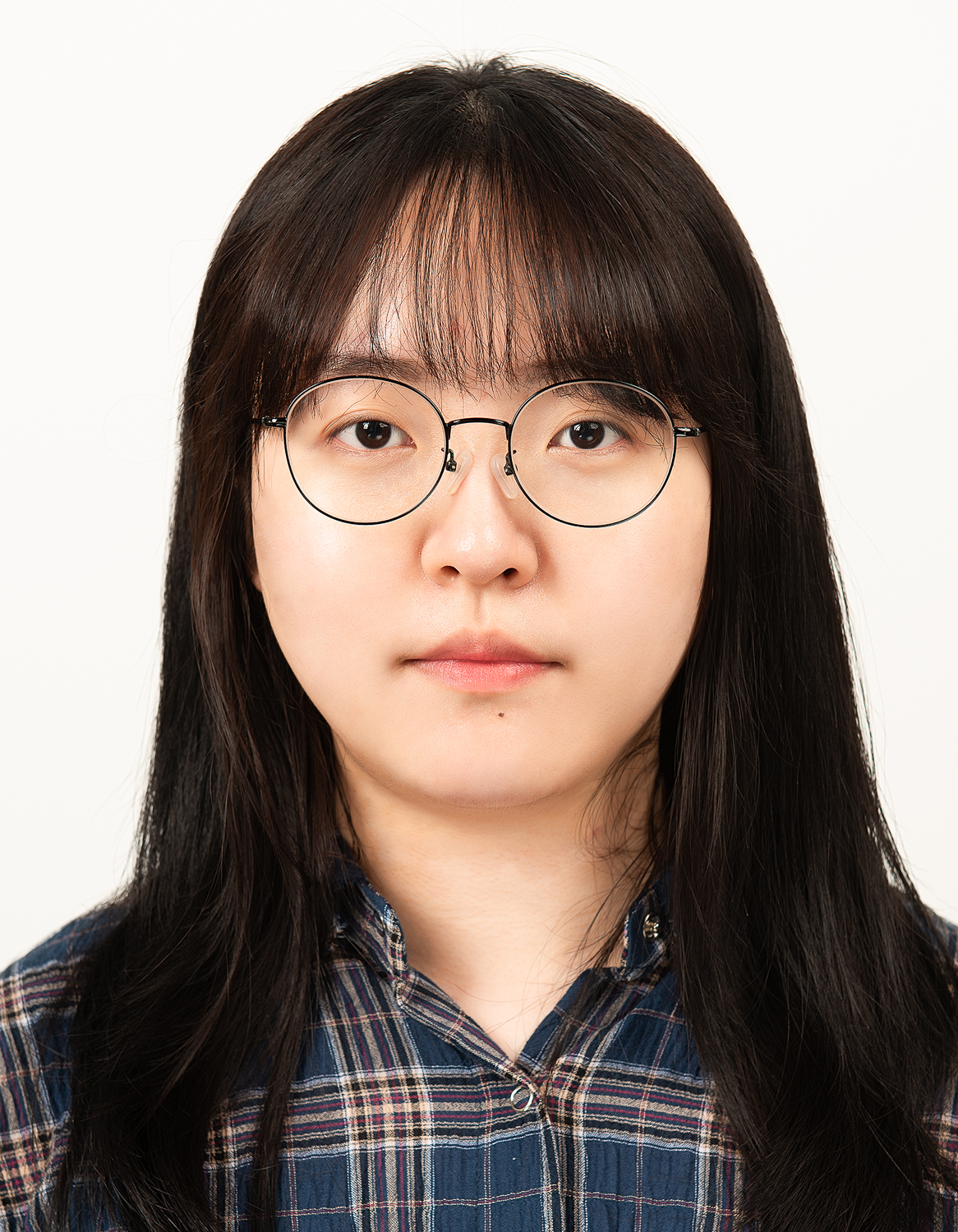}}]{Jieon Yoon}
(Graduate Student Member, IEEE) received the B.S. degree in Electrical Engineering with a double major in Computer Science from the Korea Advanced Institute of Science and Technology (KAIST), Daejeon, South Korea, in 2023, and the M.S. degree in Electrical Engineering from the same institution in 2025. She is currently pursuing a Ph.D. degree in AI Semiconductor at KAIST. Her research interests include computer architecture and design of domain-specific accelerators.
\end{IEEEbiography}

\vspace{-18pt}
\begin{IEEEbiography}
[{\includegraphics[width=1in,height=1.25in,clip,keepaspectratio]{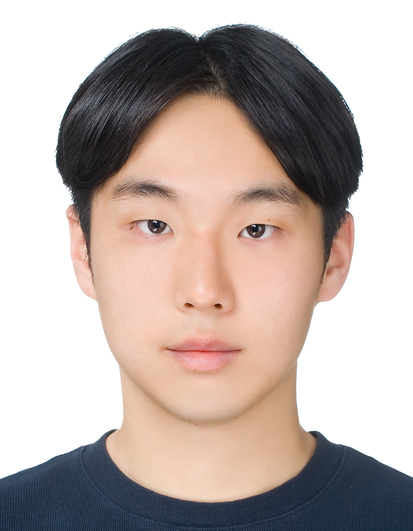}}]{Hangyeol Lee} 
(Graduate Student Member, IEEE) received his B.S. degree in Electronic Engineering from Hanyang University, Seoul, South Korea, in 2024, and his M.S. degree in Electrical Engineering from Korea Advanced Institute of Science and Technology (KAIST), Daejeon, South Korea, in 2025, where he is currently pursuing a Ph.D. His research interests include computer vision acceleration across the stack, spanning algorithmic/software optimization and software–hardware co-design for efficient model deployment.
\end{IEEEbiography}

\vspace{-18pt}
\begin{IEEEbiography}
[{\includegraphics[width=1in,height=1.25in,clip,keepaspectratio]{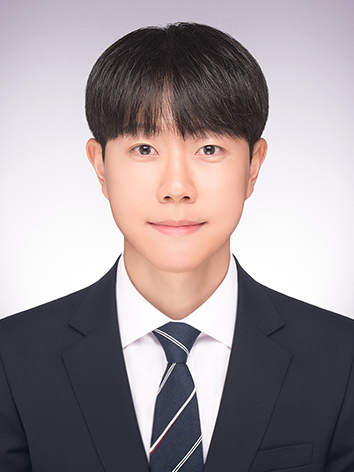}}]{Jaehoon Heo} received the B.S. degree in electronic engineering from Hanyang University, Seoul, South Korea, in 2020, and the M.S. and Ph.D. degree in electrical engineering from the Korea Advanced Institute of Science and Technology (KAIST), Daejeon, South Korea, in 2022 and 2026, respectively.

His research interests include low-power system-on-chip (SoC) design, hardware accelerators for machine learning, software–hardware co-design, and energy-efficient digital processors for deep neural network inference and training.
\end{IEEEbiography}

\vspace{-18pt}
\begin{IEEEbiography}
[{\includegraphics[width=1in,height=1.25in,clip,keepaspectratio]{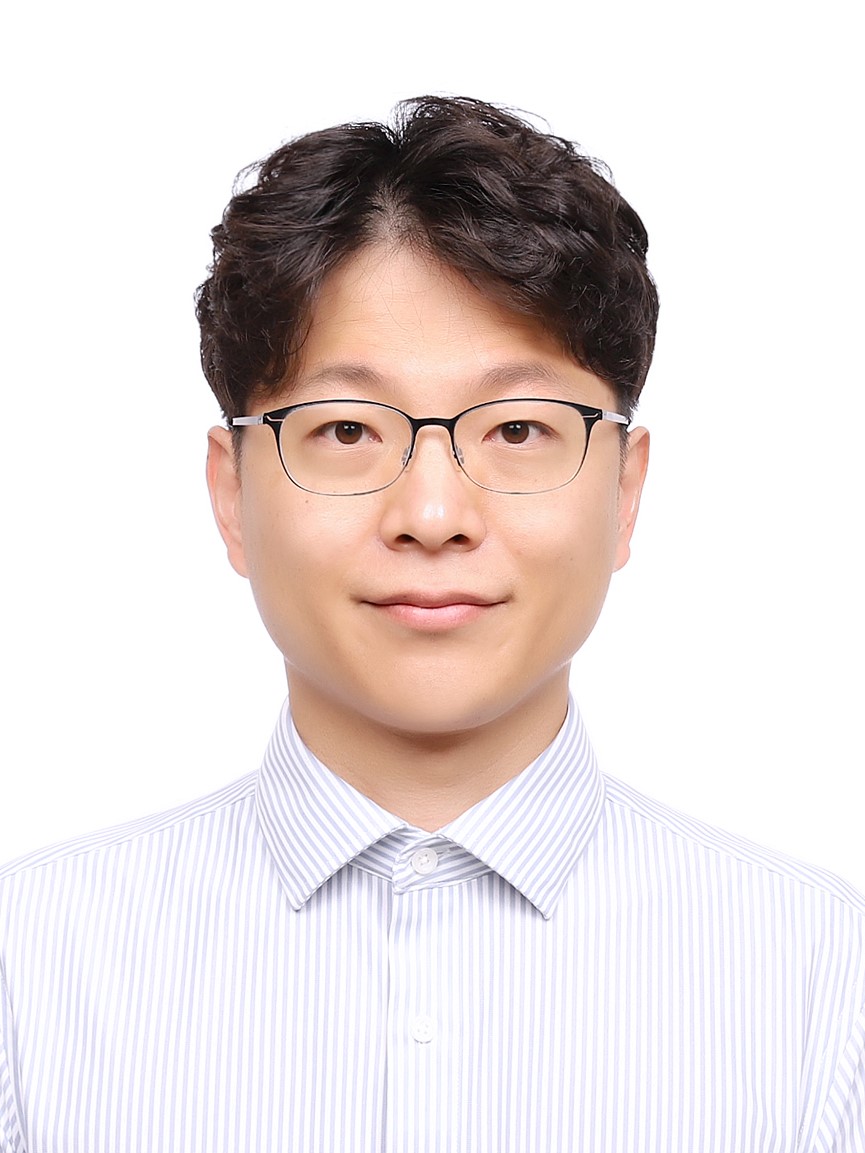}}]
{Joo-Young Kim} (Senior Member, IEEE) received the B.S., M.S., and Ph. D degrees in electrical engineering from Korea Advanced Institute of Science and Technology (KAIST), Daejeon, South Korea, in 2005, 2007, and 2010, respectively. He is currently an Associate Professor in the School of Electrical Engineering, KAIST, and the Director of the AI Semiconductor Systems Research Center, KAIST. His research interests span various aspects of hardware design, including chip design, computer architecture, domain-specific accelerators, and hardware/software co-design. Before joining KAIST, he was a Hardware Engineering Leader at Microsoft Azure, Redmond, WA, USA, working on hardware acceleration for cloud services such as machine learning, data storage, and networking.
He founded an AI fabless startup, HyperAccel, in Jan 2023 to build innovative AI processors/solutions for large-language-model(LLM)-based generative AI, making it sustainable for everyone.

\end{IEEEbiography}
\end{minipage}
}
\end{document}